\documentclass[aps,twocolumn]{revtex4}
\usepackage{amsmath}
\begin{document}
\title{Comment on ``Problems with Mannheim's conformal gravity program''}
\author{Philip D. Mannheim}
\affiliation{Department of Physics, University of Connecticut, Storrs, CT 06269, USA.
email: philip.mannheim@uconn.edu}
\date{July 24, 2015}
\begin{abstract}
Recently in Phys.~Rev.~D \textbf{88}, 027504 (2013) Yoon has suggested that there may be problems for the non-relativistic limit of the conformal gravity theory. Here we show that Yoon's results only hold because of the assumption that gravitational sources can be treated the same way that they are treated in standard Newton-Einstein gravity. Since such an assumption violates the theory's underlying conformal invariance,Yoon's conclusions are invalidated.
\end{abstract}
\maketitle

\section{Background}

In  conformal gravity with gravitational sector action $ I_{\rm W}=-\alpha_g\int d^4x (-g)^{1/2}C_{\lambda\mu\nu\kappa} C^{\lambda\mu\nu\kappa}$, one constructs a gravitational rank-two tensor $(-g)^{-1/2}\delta I_{\rm W}/\delta g_{\mu\nu}=-2\alpha_g W^{\mu\nu}$, and in the presence of a matter source with energy-momentum tensor $T^{\mu\nu}$ obtains a gravitational equation of motion of the form $-4\alpha_gW^{\mu\nu}+T^{\mu\nu}=0$. With the underlying theory being conformal invariant, both $W^{\mu\nu}$ and $T^{\mu\nu}$ are traceless. As such, conformal gravity represents two departures from standard Einstein gravity -- not only is the second-order derivative Einstein tensor $G^{\mu\nu}=R^{\mu\nu}-g^{\mu\nu}R^{\alpha}_{\phantom{\alpha}\alpha}/2$ replaced by the fourth-order derivative $W^{\mu\nu}$, the energy-momentum tensor is replaced by a traceless one. The structure of conformal gravity is thus intrinsically different from that of Einstein gravity, with experience with Einstein gravity not providing a complete enough  guide to conformal gravity.

On explicitly constructing $W^{\mu\nu}$, Mannheim and Kazanas were able to solve the conformal theory exactly in a static, spherically symmetric geometry, and showed  \cite{Mannheim1994} that the metric coefficient $B(r)=-g_{00}(r)$ obeyed the fourth-order derivative Poisson equation 
\begin{eqnarray} 
\nabla^4B(r)&=&\frac{3}{4\alpha_gB(r)}(T^0_{\phantom{0}0}-T^r_{\phantom{r}r})
=f(r).
\label{Y1}
\end{eqnarray}
\vskip-0.1cm
\noindent
Exterior to a source of radius $R$ the solution takes the form  $B(r>R)=1-2\beta/r+\gamma r$, where
\begin{eqnarray} 
2\beta=\frac{1}{6}\int_0^R dr^{\prime}f(r^{\prime})r^{\prime 4},~~\gamma=-\frac{1}{2}\int_0^R dr^{\prime}f(r^{\prime})r^{\prime 2}.
\label{Y2}
\end{eqnarray}
\vskip-0.1cm
\noindent
On finding this solution Mannheim and Kazanas realized immediately \cite{Mannheim1994} that  $\beta$ would vanish at the microscopic level if $f(r)$ was a delta function, and would vanish at the $N$ particle macroscopic level if $f(r)$ was a sum of $N$ delta functions; while, as reported in \cite{Mannheim1995}, if $f(r)$ was constant, one would then obtain $\gamma=-6\beta/ R^2$, to give a macroscopic object such as the sun an unacceptably large linear potential. However, neither option for $f(r)$ is valid in the conformal theory since if one applies $\mathbf{\nabla}^4$ to $1-2\beta/r+\gamma r$ one obtains $-8\pi\gamma \delta^3({\bf x})+8\pi\beta \mathbf{\nabla}^2 \delta^3({\bf x})$, and thus as noted in \cite{Mannheim1994,Mannheim1995}, sources must be extended or singular, with an extended $f(r)$ having both a height (cf. the $\gamma$ term) and a width (cf. the $\beta$ term), quantities that are in principle of independent strength and independent sign since they correspond to different moments of $f(r)$. However, such extendedness only needs to occur at the microscopic level, since if microscopically one obtains a potential of the form $V_p=-\beta_p/r+\gamma_p r/2$ at the level of a single proton, then for a weak gravity bulk matter source composed of $N$ protons, one can then sum over these individual microscopic sources to obtain a net $NV_p=-N\beta_p/r+N\gamma_p r/2$, just as one does in standard second-order weak gravity.

In a recent paper \cite{Yoon2013} Yoon has objected to the use of sources other than standard delta functions to thereby suggest that the coefficient $\beta$ be zero. And also Yoon has brought arguments to suggest that the coefficient $\gamma$  would be negative rather than be the positive value it has been found to possess in a successful fitting of the theory to the rotation curves of a broad class of 141 different spiral galaxies \cite{Mannheim2011,Mannheim2012,OBrien2012,Mannheim2013}. Here we refute Yoon's claims by justifying the use of extended sources, and by showing that Yoon had made an error in his analysis of the sign of $\gamma$, an error that invalidates his conclusions. 

Our refutation of both of Yoon's claims is based on recognizing that if one wants to discuss mass in a conformal theory one needs to introduce a Higgs field to generate such mass via symmetry breaking at the microscopic level. This Higgs field then carries energy density and momentum itself, and thus it is not just the energy density and momentum of the massive particles but also that of the Higgs field itself that contribute to the full energy-momentum tensor that serves as the source of gravity. Moreover, it is the interplay of the massive particles with this self-same Higgs field that then gives the massive particles their extended structure. In his paper, Yoon did not include the contribution of the Higgs field to the energy-momentum tensor (to thus invalidate his second claim). And as to his first claim, Yoon did not actually provide any analysis that might exclude the use of extended sources. Rather, his refutation consists of nothing more than his declaration that having such extended sources would be "undesirable", and in his paper  Yoon never indicated in what specific way any such extended sources might  actually be undesirable. 

Independent of the desirability or lack thereof of such extended structures, such a possible structure for sources is not familiar from standard gravity. However, it is  is also not excluded by it either, with the $-MG/r$ potential being the exact exterior solution to the second-order Poisson equation $\mathbf{\nabla}^2\phi=4\pi G\rho$ no matter how the source might behave in the interior region, In fact standard second-order gravity is insensitive to the issue since measuring the definite integral $M=4\pi \int_0^{R}dr^2r^2\rho(r)$ alone does not enable one to determine the structure or even the sign of the integrand $\rho(r)$ at points within the integration range. Since second-order gravity is not required to obey the constraints of conformal invariance, experience with second-order gravity is not a good guide as to the structure of sources in the fourth-order derivative conformal gravity case. While conformal symmetry obliges us to consider physics at the microscopic level where both masses and extended structures are generated, second-order gravity does not. Hence, central to our discussion will be the distinction between macroscopic sources and microscopic ones, and it is to this issue that we now turn.

\section{Macroscopic Considerations}

In Newtonian gravity,  the prescription for determining the macroscopic, non-relativistic potential $\phi ({\bf r})$ at any point ${\bf r}$ due to a set of $N$ static, microscopic sources of mass $m_i$ at points ${\bf r_i}$, is to sum over them according to
\begin{eqnarray}
\phi ({\bf r})= -\sum_{i=1}^{i=N} \frac{m_iG}{\vert {\bf r} -{\bf r_i},
\vert} 
\label{Y3}
\end{eqnarray}                                 
where $G$ is Newton's constant, with the motions of material test particles then being determined via
\begin{eqnarray}
\frac{d^2{\bf r}}{dt^2}=-\mathbf{\nabla}\phi.
\label{Y4}
\end{eqnarray}                                 
As such, (\ref{Y3}) and (\ref{Y4}) contain the full content of Newtonian gravity. If the microscopic sources (taken to be protons for specificity) all have the same mass $m_p$, and if they are distributed spherically symmetrically within some localized region of radius $R$, then outside of this region the potential is given by $\phi(r>R)=-Nm_pG/r$, to thus be an extensive function of the number of microscopic sources. As such, the exterior region $\phi(r>R)$ only counts the number of microscopic sources and is insensitive to their distribution within the $r<R$ region. To facilitate doing the summation over this region, one can introduce a second-order Poisson equation. To do so, one applies $\mathbf{\nabla}^2$ to (\ref{Y3}) to obtain 
\begin{eqnarray}
\mathbf{\nabla}^2\phi({\bf r})&=&4\pi Gm_p\sum_{i=1}^{i=N}  \delta^3({\bf r}-{\bf r_i})
\nonumber\\
&=&4 \pi Gm_p\sum_{i=1}^{i=N}  n({\bf r}-{\bf r_i}),
\label{Y5}
\end{eqnarray}                                 
In (\ref{Y5}) we have introduced  a number density $n({\bf r}-{\bf r_i})=\delta^3({\bf r}-{\bf r_i})$ that counts the number of microscopic sources. For sources at rest we can identify $m_pn({\bf r}-{\bf r_i})$ as the energy density $\rho({\bf r}-{\bf r_i})$.  In consequence of this the Einstein theory uses perfect fluid sources built out of such an energy density according to $T_{\mu\nu}({\rm fluid})=(\rho+p)U_{\mu}U_{\nu}+pg_{\mu\nu}$, so that its non-relativistic limit where $\rho \gg p$ recovers (\ref{Y5}). However, this is not a general rule of nature, and in the conformal case we will find a very different structure for the total $T_{\mu\nu}({\rm T})$, since the very requirement that $T_{\mu\nu}({\rm T})$ be traceless means that it cannot be comprised of the non-traceless  $T_{\mu\nu}({\rm fluid})$ alone ($T^{\mu}_{\phantom{\mu}\mu}({\rm fluid})=3p-\rho \sim -\rho \neq 0$). The extra contribution that is needed to maintain $T^{\mu}_{\phantom{\mu}\mu}({\rm T})=0$ is precisely that due to the Higgs field that gives mass to the microscopic particles in the source in the first place. Nonetheless, the counting of microscopic sources is done with the number density not the energy density, and it is such counting that conformal gravity still recovers.

For the fourth-order theory, each microscopic  proton source is to put out a potential $V_p(r)=-\beta_p/r+\gamma_p r/2$,  with the analog of (\ref{Y3})  then being given by
\begin{eqnarray}
V ({\bf r})= \sum_{i=1}^{i=N} \left[ -\frac{\beta_p}{\vert {\bf r} -{\bf r_i}\vert}  + \frac{\gamma_p\vert {\bf r} -{\bf r_i}\vert}{2}\right].
\label{Y6}
\end{eqnarray}                                 
in the non-relativistic, weak gravity regime in which each $\vert {\bf r} -{\bf r_i}\vert$ obeys $\beta_p/\vert {\bf r} -{\bf r_i}\vert \ll 1$, $\gamma_p\vert {\bf r} -{\bf r_i}\vert /2 \ll1$.
Thus again the exterior potential counts the number of microscopic sources, as per $V(r \gg R)=-N\beta_p/r+N\gamma_p r/2$. On applying $\mathbf{\nabla}^4$ to (\ref{Y6}) we obtain 
\begin{eqnarray}
\mathbf{\nabla}^4V({\bf r})&=&\sum_{i=1}^{i=N}\bigg{[}4 \pi \beta \mathbf{\nabla}^2\delta^3({\bf r}-{\bf r_i})
-4 \pi \gamma \delta^3({\bf r}-{\bf r_i})\bigg{]}.~
\label{Y7}
\end{eqnarray}                                 
As such, (\ref{Y7}) gives the $f(r)$ needed for macroscopic bulk matter in the conformal theory, as expressed in terms of microscopic potentials whose $\beta_p$ and $\gamma_p$ coefficients are not as yet constrained. Now while (\ref{Y7}) depends on the number density, just as (\ref{Y5}) does in the second-order case, the $\gamma_p$ term is not given by  the energy density since $\gamma_p$ is not $m_p$. All that one can say from macroscopic considerations is that for bulk matter both of the two summations in (\ref{Y6}) are extensive functions of the number of microscopic sources. When one uses $\rho({\bf r}-{\bf r_i})$ as the source of Einstein gravity in weak gravity bulk matter cases (the only cases where Einstein gravity has been tested), one can do so because the energy density is related to the number density of fundamental sources. For bulk applications of conformal gravity one uses (\ref{Y6}) with $\beta_p$ and $\gamma_p$ being fundamental parameters that are to be fixed from data. This has expressly been done in  \cite{Mannheim2011,Mannheim2012,OBrien2012,Mannheim2013}, where one finds that with the use of one universal value for the $\gamma^*$ of each solar mass of stars (specifically $\gamma^*=5.42\times10^{-41}~{\rm cm}^{-1}$) and one universal value for the $\beta^*$ of each solar mass of stars (viz. $\beta^*=1.48\times 10^{5}~{\rm cm}$, the standard Newtonian value), one can fit 141 galactic rotation curves without the need for any dark matter or the 282 free parameters needed for the 141 dark matter halos. With the data fitting showing that $\gamma^*$ is consistently positive, for each one of the $10^{57}$ protons in the sun one has $\gamma_p=5.42\times10^{-98}~{\rm cm}^{-1}$ (and analogously one has $\beta^*_p=1.48\times 10^{-52}~{\rm cm}$ for each proton). We thus need to see whether this positive sign for $\gamma_p$ is compatible with (\ref{Y2}) when it is  applied at the microscopic level.

\section{Microscopic Considerations}

In a theory with an underlying conformal symmetry, at the level of the Lagrangian all particles will be massless and thus propagate as unlocalized massless plane waves that fill all space. In order to give particles mass one breaks the symmetry spontaneously by giving a Higgs-type scalar field $S(x)$ a constant vacuum expectation value $S_0$. However, in and of itself this will just produce unlocalized massive plane waves that still fill all space. To get the particles to localize into some finite region of space, one needs $S(x)$ to acquire a spatial dependence in order to produce a potential well that can serve to trap and bind the particles into localized configurations. While such an extended structure description of elementary particles was not originally developed for conformal gravity per se, it nonetheless can apply to it. 

The idea of trapping a particle, typically taken to be a fermion, in a spatially-dependent field originated in studies in the 1950s of the polaron problem associated with the trapping of an electron in a polar crystal, with the some of the early relativistic models being developed in \cite{Vinciarelli1972},  \cite{Bardeen1975}, and \cite{Dashen1975}. For the illustrative double-well potential Lagrangian ${{\cal L}}=i\bar{\psi}\gamma^{\mu}\partial_{\mu}\psi+\partial_{\mu} S\partial^{\mu}/2-hS\bar{\psi}\psi -\lambda S^4+\mu^2S^2/2$ studied in \cite{Bardeen1975}, the ground state is described by a configuration in which $S(x)=S_0=\mu/2\lambda^{1/2}$. Here the constant $S_0$ corresponds to the vacuum expectation value $\langle \Omega|S(x)|\Omega\rangle$ of a quantum field $S(x)$, and since the vacuum is translation invariant, $\langle \Omega|S(x)|\Omega\rangle$ is independent of the space and time coordinates. 

Given this vacuum, it is very tempting to assume that the first excited state above this vacuum would be given by $b_{\omega}^{\dagger}|\Omega\rangle$ where $b_{\omega}$ annihilates $|\Omega\rangle$ and $b_{\omega}^{\dagger}$ is a standard fermion plane wave creation operator for a state with energy $E=\hbar\omega$ and mass $m=hS_0$. However, it turns out that there is a composite state that lies lower than this one-particle state. In it the scalar field acquires a spatially-dependent expectation value associated not with the vacuum but with a coherent state $|C\rangle$. (The state $|C\rangle$ is constructed via a spatially-dependent Bogoliubov transform on the vacuum $|\Omega \rangle$.) Now in and of itself the state  $|C\rangle$ is not an eigenstate of theory, and is not even stable. It is stabilized by the presence of the fermion so that it is $\hat{b}^{\dagger}_{\omega}|C\rangle$ that is the stable one-particle eigenstate where $\hat{b}_{\omega}$ annihilates $|C\rangle$, and where the spatial part of the fermion wave function is now localized and is no longer a momentum eigenstate. In this configuration $S(x)$ is given by  $\langle C|\hat{b}_{\omega}S({\bf r})\hat{b}^{\dagger}_{\omega}|C\rangle$, and the fermion localizes in this spatially-dependent potential when it is inserted into the fermionic Dirac equation. The fermion and the coherent state thus mutually stabilize each other into a  localized, so-called bag model, extended structure configuration, with it being the bag pressure due to the spatially-dependent scalar field that both stabilizes and localizes the fermion. In \cite{Eguchi1974}, \cite{Dashen1975}, and \cite{Mannheim1976} these ideas have been extended to the case where the symmetry breaking is done by the fermion composite operator $\bar{\psi}\psi$. And in \cite{Mannheim1978} these ideas have been extended to a conformal invariant formulation of quantum electrodynamics (in flat spacetime), where it was shown that the bag pressure could serve as the Poincare stresses associated with a completely electrodynamical electron mass. In all of these extended structure pictures there is a bag pressure, and thus when the theory is coupled to gravity the bag pressure will become part of the gravitational source.

To discuss the gravitational situation, rather than consider the full bag model developed in \cite{Mannheim1978} and recently revisited in \cite{Mannheim2015}, it is simpler to consider a model consisting of a fermion conformally coupled to a fundamental scalar field \cite{Mannheim2006} as the model is rich enough to illustrate the issues that are involved. For this model the  Lagrangian is given by ${{\cal L}}=i\bar{\psi}\gamma^{\mu}(x)[\partial_{\mu}+\Gamma_{\mu}(x)]\psi +\partial_{\mu} S\partial^{\mu}S/2-S^2R^{\mu}_{\phantom{\mu}\mu}/12+\lambda S^4-hS\bar{\psi}\psi $. (Here $\gamma^{\mu}(x)$ is a spacetime-dependent Dirac gamma matrix and  $\Gamma_{\mu}(x)$ is the fermion spin connection). The scalar field and fermion wave equations are given by
\begin{eqnarray}
&&\nabla_{\mu}\nabla^{\mu}S+SR^{\mu}_{\phantom{\mu}\mu}/6-4\lambda S^3+h\bar{\psi}\psi=0,
\nonumber\\                                                                              
&&i\gamma^{\mu}(x)[\partial_{\mu}+\Gamma_{\mu}(x))]\psi -hS\psi=0.               
\label{Y8}
\end{eqnarray}
The total $T_{\mu\nu}({\rm T})$ is given by  $T_{\mu\nu}({\rm T})=T_{\mu\nu}({\rm F})+T_{\mu\nu}({\rm S})$, where 
\begin{eqnarray}                                                                              
&&\!\!T_{\mu\nu}(\rm S)=2\nabla_{\mu}S \nabla_{\nu}S /3 -g_{\mu\nu}\nabla^{\alpha}S\nabla_{\alpha}S/6 -S\nabla_{\mu}\nabla_{\nu}S/3
\nonumber\\
&&\!\!+g_{\mu\nu}S\nabla^{\alpha}\nabla_{\alpha}S/3 -S^2(2R_{\mu\nu}-g_{\mu\nu}R^\alpha_{\phantom{\alpha}\alpha})/12
-g_{\mu\nu}\lambda S^4,
\nonumber\\
&&\!\!T_{\mu\nu}({\rm F})=i\bar{\psi}\gamma_{\mu}(x)[\partial_{\nu}+\Gamma_{\nu}(x)]\psi.             
\label{Y9}
\end{eqnarray}

In the presence of a radially-dependent Higgs scalar field $S(r)$ and a fermion that is bound to it, the radial and angular sectors of the theory have very different $r$ dependencies. Hence neither $T_{\mu\nu}({\rm F})$ nor $T_{\mu\nu}({\rm S})$ can take the form of a perfect fluid since for a perfect fluid one must have $T^{r}_{\phantom{r}r}=T^{\theta}_{\phantom{\theta}\theta}$. However, in a static spherically symmetric geometry both tensors are diagonal. Thus for  $T_{\mu\nu}({\rm F})$ we shall set $T^{0}_{\phantom{0}0}({\rm F})=-R(r)$, $T^{r}_{\phantom{r}r}({\rm F})=P(r)$, $T^{\theta}_{\phantom{\theta}\theta}({\rm F})=T^{\phi}_{\phantom{\phi}\phi}({\rm F})=P(r)+Q(r)$ \cite{footnote1}. In terms of these quantities we thus obtain  (see e.g. \cite{Mannheim2007})
\begin{eqnarray}                                                                              
f_{\rm F}(r)&=& \frac{1}{4\alpha_g B}\left[-3(R+P)
+BSS^{\prime \prime}-2BS^{\prime 2}\right],                 
\nonumber\\
T^{\mu}_{\phantom{\mu}\mu}(\rm T)&=& \frac{S^2}{6r^2}\left[r^2B^{\prime\prime}+4rB^{\prime}+2B-2\right] +3P+2Q-R
\nonumber\\
&+&\frac{S}{r}\left[rBS^{\prime\prime}+2BS^{\prime}+rB^{\prime}S^{\prime}\right]-4\lambda S^4 =0,
\label{Y10}
\end{eqnarray}
where $f_{\rm F}(r)$ denotes the $f(r)$ of a single fermion. In terms of this $f_{\rm F}(r)$ the macroscopic $f(r)$ needed for (\ref{Y7}) is given by $f(r)=\sum_i f_{\rm F}({\bf r_i})\delta^3({\bf r}-{\bf r_i})$, with the number operator doing the counting just as in the Newtonian case.

Now in his paper Yoon tries to fix the sign of $\gamma_p$ by taking $f(r)$ to be dominated by $R(r)$ alone. However from the vanishing trace condition for $T_{\mu\nu}({\rm T})$, we see that the other terms in $T^{\mu}_{\phantom{\mu}\mu}(\rm T)$ are necessarily just as big. Since Yoon does dominate $f(r)$ by $R(r)$, and since he does not take into account the contribution of the Higgs field to $T_{\mu\nu}({\rm T})$, his conclusions regarding the sign of $\gamma$ are not valid \cite{footnote2}.


\begin{thebibliography}{99}
%
\bibitem{Mannheim1994} P.~D.~Mannheim~and~D~Kazanas,~Gen.~Rel.~Gravit. \textbf{26}, 337 (1994).
%
\bibitem{Mannheim1995} P.~D.~Mannheim, \textit{Four dimensional conformal gravity, confinement, and 
galactic rotation curves}, in Proceedings of ``PASCOS 94", the Fourth 
International Symposium on Particles, Strings and Cosmology, Syracuse, New York.  K. C. Wali Ed. (World Scientific Press, Singapore, 1995). 
%
\bibitem{Yoon2013}  Y.~Yoon,~Phys.~Rev.~D \textbf{88}, 027504 (2013).
%
\bibitem{Mannheim2011} P.~D.~Mannheim~and~J.~G.~O'Brien, Phys.~Rev.~Lett. \textbf{106}, 121101 (2011).
%
\bibitem{Mannheim2012} P.~D.~Mannheim~and~J.~G.~O'Brien,~Phys.~Rev.~D \textbf{85}, 124020 (2012).
%
\bibitem{OBrien2012} J.~G.~O'Brien~and~P.~D.~Mannheim, Mon.~Not.~R.~Astron.~Soc. \textbf{421}, 1273 (2012).
%
\bibitem{Mannheim2013} P.~D.~Mannheim~and~J.~G.~O'Brien,~J.~Phys.~Conf.~Ser. \textbf{437}, 012002 (2013). 
%
\bibitem{Vinciarelli1972} P.~Vinciarelli,~Nouvo.~Cim.~Lett~\textbf{4}, 905 (1972).
%
\bibitem{Bardeen1975} W.~A.~Bardeen,~M.~S.~Chanowitz,~S.~D.~Drell,~M.~Weinstein, and~T.-M.~Yan, Phys. Rev. D \textbf{11}, 1094 (1975).
%
\bibitem{Dashen1975} R.~F.~Dashen,~B.~Hasslacher,~and~A.~Neveu,~Phys.~Rev.~D \textbf{12}, 2443 (1975).
%
\bibitem{Eguchi1974} T.~Eguchi~and~H.~Sugawara,~Phys.~Rev.~D~\textbf{10}, 4257 (1974).
%
\bibitem{Mannheim1976}  P.~D.~Mannheim,~Phys.~Rev.~D~\textbf{14}, 2072 (1976).
%
\bibitem{Mannheim1978} P.~D.~Mannheim,~Nucl.~Phys.~B~\textbf{143}, 285 (1978).
%
\bibitem{Mannheim2015} P.~D.~Mannheim, \textit{Living without supersymmetry -- the conformal alternative and a dynamical Higgs boson}, arXiv:1506.01399 [hep-ph], June, 2015.
%
\bibitem{Mannheim2006} P.~D.~Mannheim,~Prog. Part.~Nucl.~Phys. \textbf{56}, 340 (2006).
%
\bibitem{footnote1} With the papers of Mannheim and collaborators being written with a metric in which $g_{00}$ is negative, then as per the sign conventions for perfect fluids  of the generic form $T_{\mu\nu}=(\rho+p)U_{\mu}U_{\nu}+pg_{\mu\nu}$,  $T^{\mu}_{\phantom{\mu}\mu}=3p-\rho$,  $g^{\mu\nu}U_{\mu}U_{\nu}=-1$, we have $T_{00}({\rm F})=BR$, $T^{0}_{\phantom{0}0}({\rm F})=-R$. 
%
\bibitem{Mannheim2007} P.~D.~Mannheim,~Phys.~Rev.~D \textbf{75}, 124006 (2007).
%
\bibitem{footnote2} In his paper Yoon makes two other errors, errors which actually compensate each other. First, he  sets $T^{0}_{\phantom{0}0}({\rm F})=+R$. However, in \cite{footnote1}  we  noted that $T^{0}_{\phantom{0}0}({\rm F})=-R$. Second, Yoon asserts without proof that the sign of the coupling constant $\alpha_g$ is positive. However, in P.~D.~Mannheim, ~Gen.~Rel.~Gravit. \textbf{43}, 703 (2011) it was found that it is a negative sign for $\alpha_g$ that leads to a positive sign for the zero-point energy of the bosonic gravitational field, just as is needed to cancel the negative zero-point energy density of the fermion field. Also one can show (P.~D.~Mannheim, unpublished) that it is a negative $\alpha_g$ that makes the Euclidean time conformal gravity action  be negative on every path integral path, in consequence of which the energy spectrum of the gravitational sector of quantum conformal gravity is bounded from below, just as is needed for a viable quantum theory.
%



\end{thebibliography}
\end{document}